# A DATA MINING APPROACH TO PREDICT PROSPECTIVE BUSINESS SECTORS FOR LENDING IN RETAIL BANKING USING DECISION TREE

Md. Rafiqul Islam[1] and Md. Ahsan Habib[2]

Department of Information and Communication Technology
Mawlana Bhashani Science and Technology University, Tangail, Bangladesh

***ABSTRACT***

*A potential objective of every financial organization is to retain existing customers and attain new prospective customers for long-term. The economic behaviour of customer and the nature of the organization are controlled by a prescribed form called Know Your Customer (KYC) in manual banking. Depositor customers in some sectors (business of Jewellery/Gold, Arms, Money exchanger etc) are with high risk; whereas in some sectors (Transport Operators, Auto-delear, religious) are with medium risk; and in remaining sectors (Retail, Corporate, Service, Farmer etc) belongs to low risk. Presently, credit risk for counterparty can be broadly categorized under quantitative and qualitative factors. Although there are many existing systems on customer retention as well as customer attrition systems in bank, these rigorous methods suffers clear and defined approach to disburse loan in business sector. In the paper, we have used records of business customers of a retail commercial bank in the city including rural and urban area of (Tangail city) Bangladesh to analyse the major transactional determinants of customers and predicting of a model for prospective sectors in retail bank. To achieve this, data mining approach is adopted for analysing the challenging issues, where pruned decision tree classification technique has been used to develop the model and finally tested its performance with Weka result. Moreover, this paper attempts to build up a model to predict prospective business sectors in retail banking.*

## 1. INTRODUCTION

Rising numbers of financial institutions are introducing and expanding their offerings of electronic banking products. Central to the business strategy of every financial service company is the ability to retain existing customer and reach new prospective customers. The Banking industry in Bangladesh is growing rapidly and it has become more and more important to keep pace with the growth of the industry through technological advancements and innovative ideas to market the organization to the masses. Portfolio of products offered by bank providers has diversified, over the years, attracting more customers than ever. Accumulation of operational data inevitably follows from this growth in industry. There exists an increasing need to convert their data into a corporate asset in order to stay ahead and gain a competitive advantage. Data mining is adopted to play an important role in these efforts. Data mining is an iterative process that combines business knowledge, machine learning methods and tools and large amounts of accurate and relevant information to enable the discovery of non-intuitive insights hidden in the organization's corporate data. This information can refine existing processes, uncover trends and help formulating policies regarding the company's relation to its customers and employees. In the financial area, data mining has been applied successfully in determining the likely eligible candidate for loan disbursement, finding profitable customers, products, characterizing different product segments [1]. All of these factors are challenging old ways of doing business and forcing banks to consider reinventing themselves to win in the marketplace. In this aspect to find out good customers to disbursing loan is really a challenging issue in the banking era. This paper is trying to find out the prospective business sectors for retail banking.

## 2. RELATED WORKS

Several researches have been made in the field of customer attrition and retention analysis in banking sectors. Some studies reveal that the most important variables influencing customer choice are effective and efficient customer services, speed and quality services, variety of services offered and low e-service charges, online banking facilities, safety of funds and the availability of technology based service(s), low interest rate on loan, convenient branch location, image of the bank, well management, and overall bank environment [7-9]. On the other hand, customer is the core of their operation, so nurturing and retaining them are important for their success. Many researches were held on customer retention as well as customer attrition analysis Lift is used as a proper measure for attrition analysis and compare the lift of data mining models of decision tree, boosted naive Bayesian network, selective Bayesian network, neural network and the ensemble of classifiers of the above methods [1]. Their main focuses were on attrition analysis using lift. Lift can be calculated by looking at the cumulative targets captured up to p% as a percentage of all targets and dividing by p%. A churn model with a higher predictive performance in a newspaper subscription context was constructed support vector machines [10]. They showed that support vector machines show good generalization performance when applied to noisy marketing data. The model outperforms a logistic regression only when the appropriate parameter-selection technique is applied and SVMs are surpassed by the random forests. A software using Clementine was used to analyzed 300 records of customers Iran Insurance Company in the city of Anzali, Iran [3]. They used demographic variables to determine the optimal number of clusters in K-means clustering and evaluated binary classification methods (decision tree QUEST, decision tree C5.0, decision tree CHAID, decision trees CART, Bayesian networks, Neural networks) to predict customers churn [3, and 5]. "Ref. [6]" used Decision trees and Neural Networks to develop model to predict churn. Models generated are evaluated using ROC curves and AUC values. They also adopted cost sensitive learning strategies to address imbalanced class labels and unequal misclassification costs issues. "Ref. [4]" discussed commercial bank customer churn prediction based on SVM model, and used random sampling method to improve SVM model, considering the imbalance characteristics of customer data sets. A study investigated determinants of customer churn in the Korean mobile telecommunications service market based on customer transaction and billing data. Their study defines changes in a customer's status from active use to non-use or suspended as partial defection and from active use to churn as total defection. Results indicate that a customer's status change explains the relationship between churn determinants and the probability of Churn [15]. A neural network (NN) based approach to predict customer churn in subscription of cellular wireless services. Their results of experiments indicate that neural network based approach can predict customer churn with accuracy more than 92% [11]. An academic database of literature between the periods of 2000–2006 covering 24 journals and proposes a classification scheme to classify the articles. Nine hundred articles were identified and reviewed for their direct relevance to applying data mining techniques to CRM. They found that the research area of customer retention received most research attention; and classification and association models are the two commonly used models for data mining in CRM [2]. A critique on the concept of Data mining and Customer Relationship Management in organized Banking and Retail industries was also discussed [14].

Most of these paper used existing customer's data from a single database. Some of them used only demographic data. But in our system, we used data from different branches of a bank and merge these into a single database. We have analyzed borrower's transactional data. We focused on predicting prospective business sectors to disburse loan in retailing commercial bank.

## 3. PROPOSED SYSTEM

The proposed system is shown in "Figure 1". Customer transactional related data are taken from database and data mining pre-processing techniques are applied to these data. After some statistical analyzing, a pre-defined hierarchical class (target value or voluntary attributes) are assigned on per account data. These data are then fit into a classification method. A decision tree classification method is developed to extract rules to predict prospective sectors.

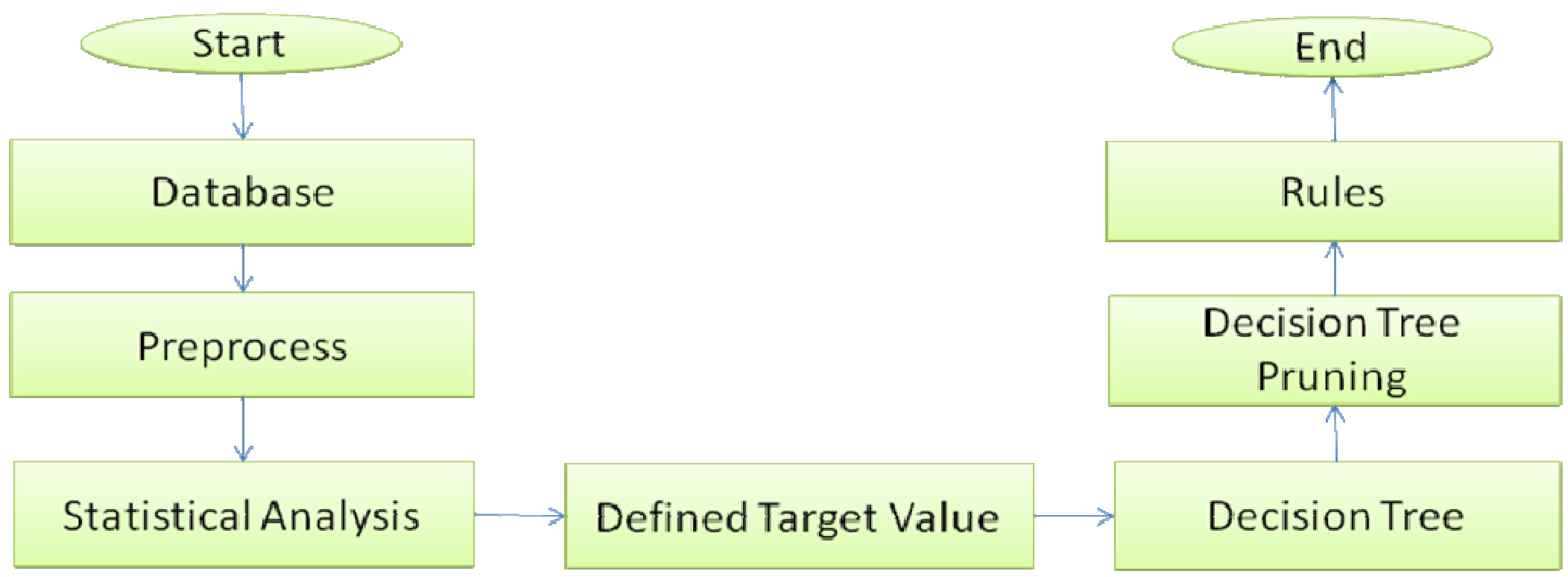

Figure 1. Overview of proposed system

## 4. SYSTEM DESIGN AND IMPLEMENTATION

### 4.1. Data period identification

Given the data availability and its time periodicity, decided to start the data selection by extracting a subset of records of accounts whose status was "Active" for the period of six month from loan disburse or reschedule date which is not later to 31/12/2013. Due to one year period of business loan, 18 months of transactional records were obtained from database. This means, records were selected not later to 30/06/2014 as well as not prior to 01/01/2013. "fromDate" and "toDate" is a subquery used in main query to retrieve data from database. For comparison purposes, only transactional data were requested whose transaction type code was either in (1, 2, 3, 4, 11, 12, 13, 18, 21, 22, 23, 24, 25, 26, 27, 29, 30) where 1 for CASH, 2 for CLEARING, 11 for PAY ORDER, 29 for ALL OTHER CREDIT TRANSACTIONS, 30 for ALL OTHER DEBIT TRANSACTIONS etc. Based on this consideration, a random number of records were selected. The records obtained are stored in a temporary table to be used to join with all relevant tables fetched in subsequent sections.

### 4.2. Data selection

The data extracted from the data source had one row per account for basic data such as account number, account title, sector, loan disbursement authority etc. The result of basic data is shown in "Table 1". The values of these static variables do not change over time. The values of time sensitive variables mean transactional data changes from month to month and it is essential to retain all these different values for 6 months in order to find seasonal/temporal behaviour related to the account activity. Transactional data i.e. minimum credit amount of single transaction of each account was fetched from database using the following query where "fromDate" and "toDate" were described in previous section. The query ignored system generated transactions such as Incidental Charges, Closing Charges, Service Charges, Interest, Source Taxes on Interest, Postage, Inspection Charges, Other Charges, Value Added Tax, Commission, Insurance premium,

Error Correction, and Miscellaneous Adjustment. The output of the query was examined in SELECT for null value and if found then replaced them with ZERO.

Table 1. Customer's account-wise basic static data

| AccountNo | Sector | Sanction Authority | Limit |
|---|---|---|---|
| 2011000851 | Other | RO | 300000 |
| 2011000745 | RiceandFlowers | DO | 3500000 |
| 2011000174 | RiceandFlowers | RO | 800000 |
| 2011000893 | Wholeseller | RO | 1600000 |
| 2011000877 | BusinessmanInd. | RO | 1000000 |
| --- | --- | --- | --- |

## 4.3. Statistical Analysis

A significant portion of the time-series data set is constants, and null fields. Filtering these fields out in the early stage can significantly reduce the processing time and improve the model accuracy [1]. The 11% of records were a single valued attributes and 9% of records were a null valued, was deemed statistically insignificant. These fields are removed from the target data table to reduce the computing time for ensuring the modelling process faster. Business customer's account wise transactional numerical continuous data is shown in "Figure 2".

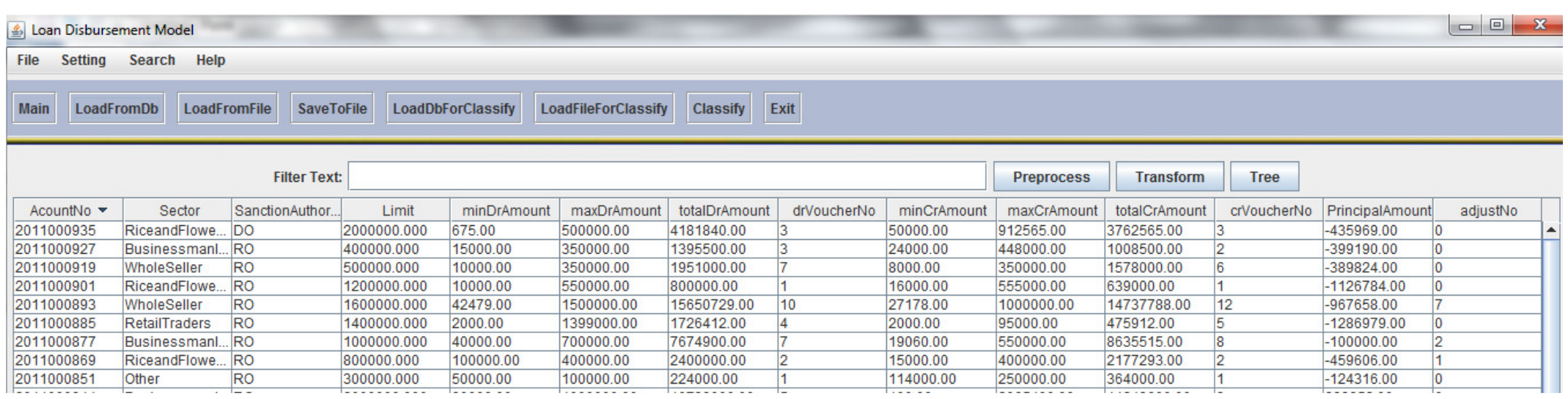

| AcountNo | Sector | SanctionAuthor... | Limit | minDrAmount | maxDrAmount | totalDrAmount | drVoucherNo | minCrAmount | maxCrAmount | totalCrAmount | crVoucherNo | PrincipalAmount | adjustNo |
|---|---|---|---|---|---|---|---|---|---|---|---|---|---|
| 2011000935 | RiceandFlowe... | DO | 2000000.000 | 675.00 | 500000.00 | 4181840.00 | 3 | 50000.00 | 912565.00 | 3762565.00 | 3 | -435969.00 | 0 |
| 2011000927 | Businessmanl... | RO | 400000.000 | 15000.00 | 350000.00 | 1395500.00 | 3 | 24000.00 | 448000.00 | 1008500.00 | 2 | -399190.00 | 0 |
| 2011000919 | WholeSeller | RO | 500000.000 | 10000.00 | 350000.00 | 1951000.00 | 7 | 8000.00 | 350000.00 | 1578000.00 | 6 | -389824.00 | 0 |
| 2011000901 | RiceandFlowe... | RO | 1200000.000 | 10000.00 | 550000.00 | 800000.00 | 1 | 16000.00 | 555000.00 | 639000.00 | 1 | -1126784.00 | 0 |
| 2011000893 | WholeSeller | RO | 1600000.000 | 42479.00 | 1500000.00 | 15650729.00 | 10 | 27178.00 | 1000000.00 | 14737788.00 | 12 | -967658.00 | 7 |
| 2011000885 | RetailTraders | RO | 1400000.000 | 2000.00 | 1399000.00 | 1726412.00 | 4 | 2000.00 | 95000.00 | 475912.00 | 5 | -1286979.00 | 0 |
| 2011000877 | Businessmanl... | RO | 1000000.000 | 40000.00 | 700000.00 | 7674900.00 | 7 | 19060.00 | 550000.00 | 8635515.00 | 8 | -100000.00 | 2 |
| 2011000869 | RiceandFlowe... | RO | 800000.000 | 100000.00 | 400000.00 | 2400000.00 | 2 | 15000.00 | 400000.00 | 2177293.00 | 2 | -459606.00 | 1 |
| 2011000851 | Other | RO | 300000.000 | 50000.00 | 100000.00 | 224000.00 | 1 | 114000.00 | 250000.00 | 364000.00 | 1 | -124316.00 | 0 |

Figure 2. Account wise transactional numerical continuous data.

## 4.4. Data transformation

Most of the data in the final table were numerical as the main concerned of the paper was on transactional data of business customers. All of the researcher used min-max, z-score or decimal scaling normalization. We used turnover to normalize the numerical data defined in equation 01.

$$\text{Turnover(A)} = (\text{Attribute}/\text{SanctionAmount})*100 \qquad \text{---} \qquad (01)$$

The turnovers of the attributes (Minimum Credit Amount, Maximum Credit Amount, Minimum Debit Amount, and Maximum Debit Amount) were clustered into four groups LessEqual25, LessEqual50, LessEqual75, and Above75. Whereas the turnovers of the attributes (Total Debit Amount, Total Credit Amount) were clustered into six groups LessEqual100, LessEqual200, LessEqual300, LessEqual400, LessEqual500, and Above500. The turnover of the attribute (Total Credit Voucher) was clustered into three groups Exceed Limit, LessEqual50, and Above50. The monthly average numbers of the attributes (Total Debit Voucher, Total Credit Voucher) were clustered into four groups LessEqual3, LessEqual6, LessEqual10, and Above10. Finally, the Loan Adjustment attribute was clustered into two groups Adjusted, and NoAdjusted.

## 4.5. Target value definition

It is part of the data mining procedure to define the proper target field based on the business objective for the data mining analysis. With the help of the business domain experts who were involved in loan department, define the target value in terms of existing data and, with these, define the value of the target variable, i.e. the variable that determines the voluntary attributes, hereby defined as Class_Label is shown in “Figure 3”. It is defined in terms of: Excellent (Equal and Above 90); Very Good (Equal and Above 80 but Less Than 90); Good (Equal and Above 70 but Less Than 80); Marginal (Equal and Above 60 but Less Than 70); and Bad (Less Than 60).

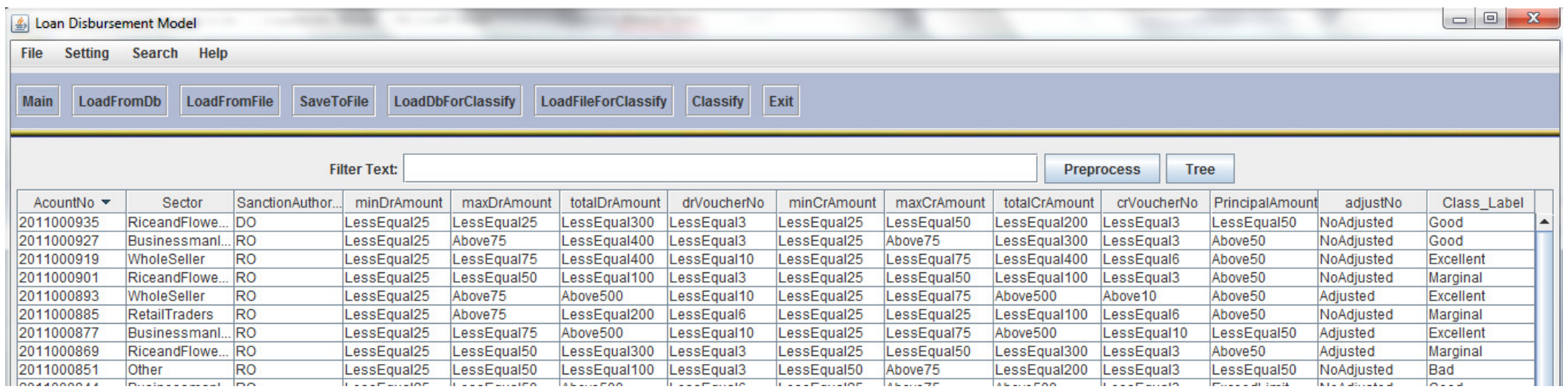

Figure 3. Account wise transactional discretized data.

## 4.6. Decision Tree

Decision tree induction is the learning of decision trees from class-labelled training tuples. A decision tree is a flowchart-like tree structure which builds a collection of rules for use as a predictive model. The advantage of this approach is that the rules are easy to understand, and they are frequently useful for discovering underlying business processes. Most algorithms such as ID3, C4.5, and CART for decision tree induction follow a greedy, top-down recursive divide-and-conquer approach, which starts with a training set of tuples and their associated class labels. The training set is recursively partitioned into smaller subsets as the tree is being built. Jiawei Han et al in their book described a basic decision tree algorithm [16]. Summary of the algorithm is as follows:

1. Basic algorithm (a greedy algorithm)
   - A top-down recursive divide-and-conquer manner
   - Start as a single node, representing all the training examples are at the root
   - Attributes are discretized for continuous-valued
   - The algorithm uses the same process recursively to partition tuples based on selected attributes
   - At each node, choose the attribute with the largest information gain
2. Conditions for stopping partitioning
   - All tuples for a given node belong to the same class
   - There are no remaining attributes for further partitioning – majority voting is adopted for classifying the leaf
   - There are no tuples left for a given node

We used specialized form of decision tree algorithm [16], which starts with a training set of tuples and their associated class labels and additional parameter arc (also known as an edge from parent to child). The initial value of arc is null since root node of the tree has no parent. So parameters of method look like generate( D, attributeList , parentPath) and little changed in the algorithm. For attribute selection, we used gain ratio which attempts to overcome unique identifier. We picked up capital letter from the descritized attribute (and added prefix max or min in appropriate cases) to graphically represent these attributes in the tree is shown in ‘Figure 04”. For example SA means Sanction Authority, maxCA means maximum Credit Amount, G means Good, VG means Very Good, etc.

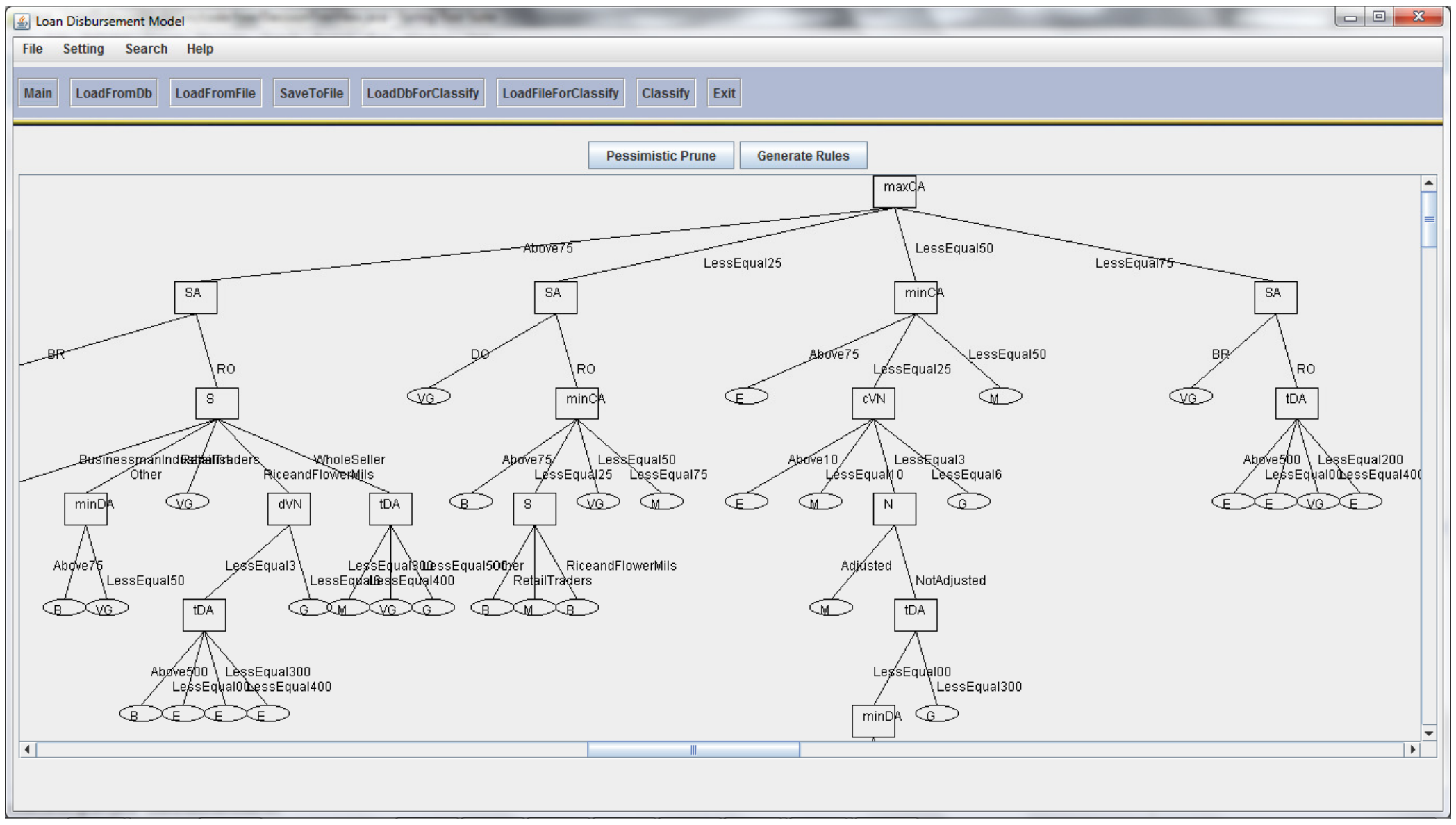

Figure 4. Partial output of the decision tree algorithm

### 4.7 Decision Tree Pruning

Tree pruning methods convert a large tree into a small tree, making it easier to understand, to improve accuracy and prevent "overfitting". The empirical comparisons of different pruning methods across several domains were examined by several researchers [17, 18]. In the paper, we used pessimistic tree pruning algorithm which avoid the necessity of a separate test data set. We first set the number of correctly or mis-correctly classified tuples at each leaf node based on training tuple. Then the following algorithm is used to prune the tree and the output of pruned tree shown in Figure 5.

//Input: Node is the root node of the tree.
//Output: Return the pruned version of the tree.
void Prune(node){
Step 1. If node is leaf then return;
Step 2. Find the majority class in sub-tree (in this case node) and set it to *MajorityClass*
Step 3. Calculate the number of mis-correctly classified tuples at each leaf with *MajorityClass* and set it to a variable *noMisLeaf*
*noMisLeaf* = *noMisLeaf* + 0.5 (1/2 for single node i.e leaf);
Step 4. Calculate the number of mis-correctly classified tuples at node (sub-tree) and set it to *noMisSubTree*.
*noMisSubTree* = *noMisSubTree* + $N_T$/2.0; (where $N_T$ is total number of leaf in sub-tree)
Step 5. Set *Total* = *noMisLeaf* + *noMisSubTree*;
Step 6. Calculate standard error using
*StandardError* = Math.sqrt((*noMisSubTree* *( *Total* - *noMisSubTree*))/*Total*);
Step 7. Check for pruning condition
IF *noMisSubTree*+ *StandardError* >= *noMisLeaf* THEN
//Pruned the tree i.e replaced sub-tree with leaf.
Set node level to *MajorityClass*, node child to null (means convert it to leaf), add *noMisLeaf*, to node and return;

Step 8. Iterate over each child of node
    Prune(child)
Step 10. Return.

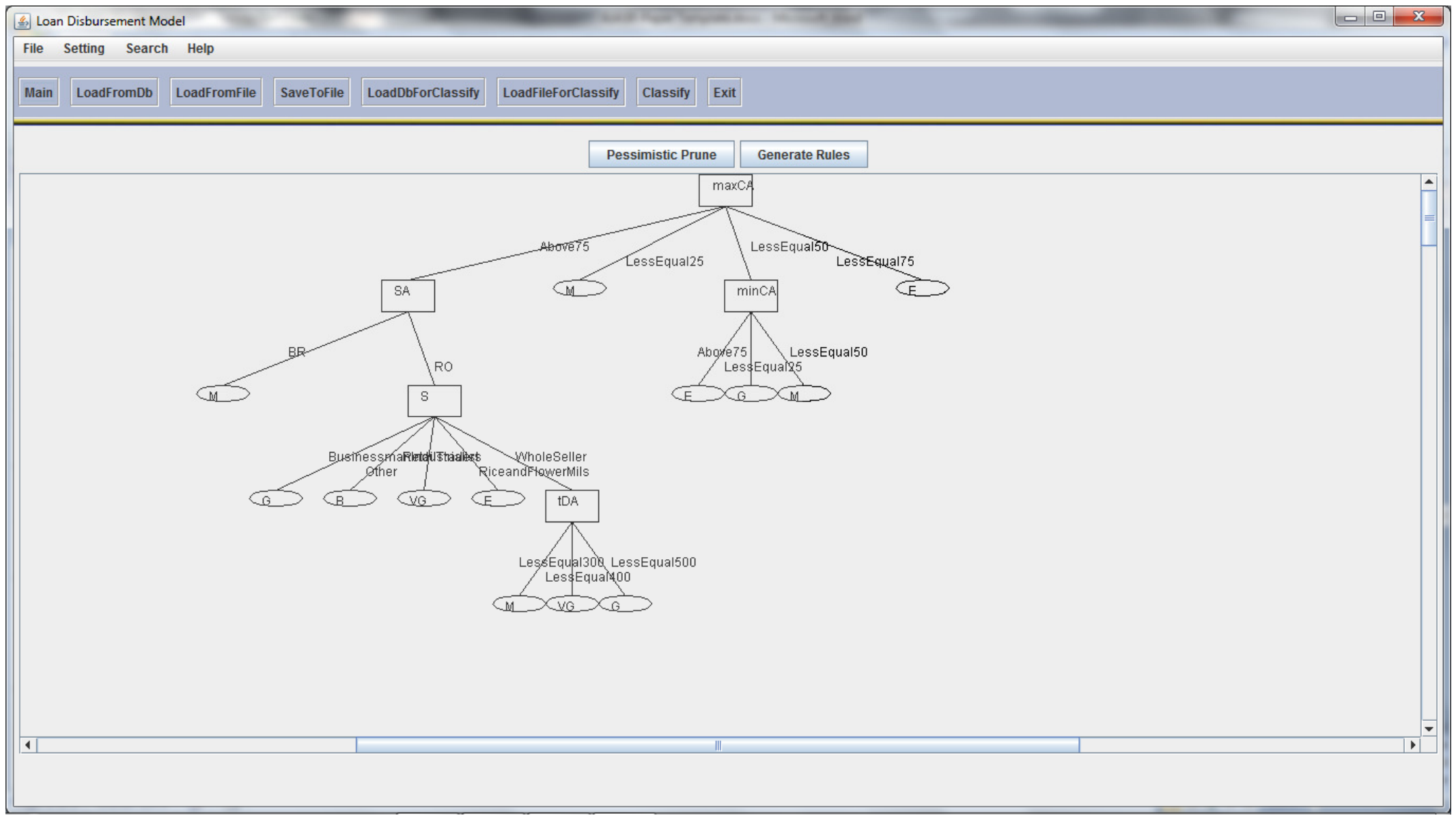

Figure 5. Output of pruned tree

## 4.8. Rules Extraction from Tree

If the decision tree is very large, then the IF-THEN rules are easier for humans to understand. To extract rules from a decision tree, one rule is created for each path from the root to a leaf node. Each splitting criterion along a given path is logically ANDed to form the rule antecedent ("IF" part). The leaf node holds the class prediction, forming the rule consequent ("THEN" part) [16]. Some generated rules from pruned tree are shown in "Figure 6".

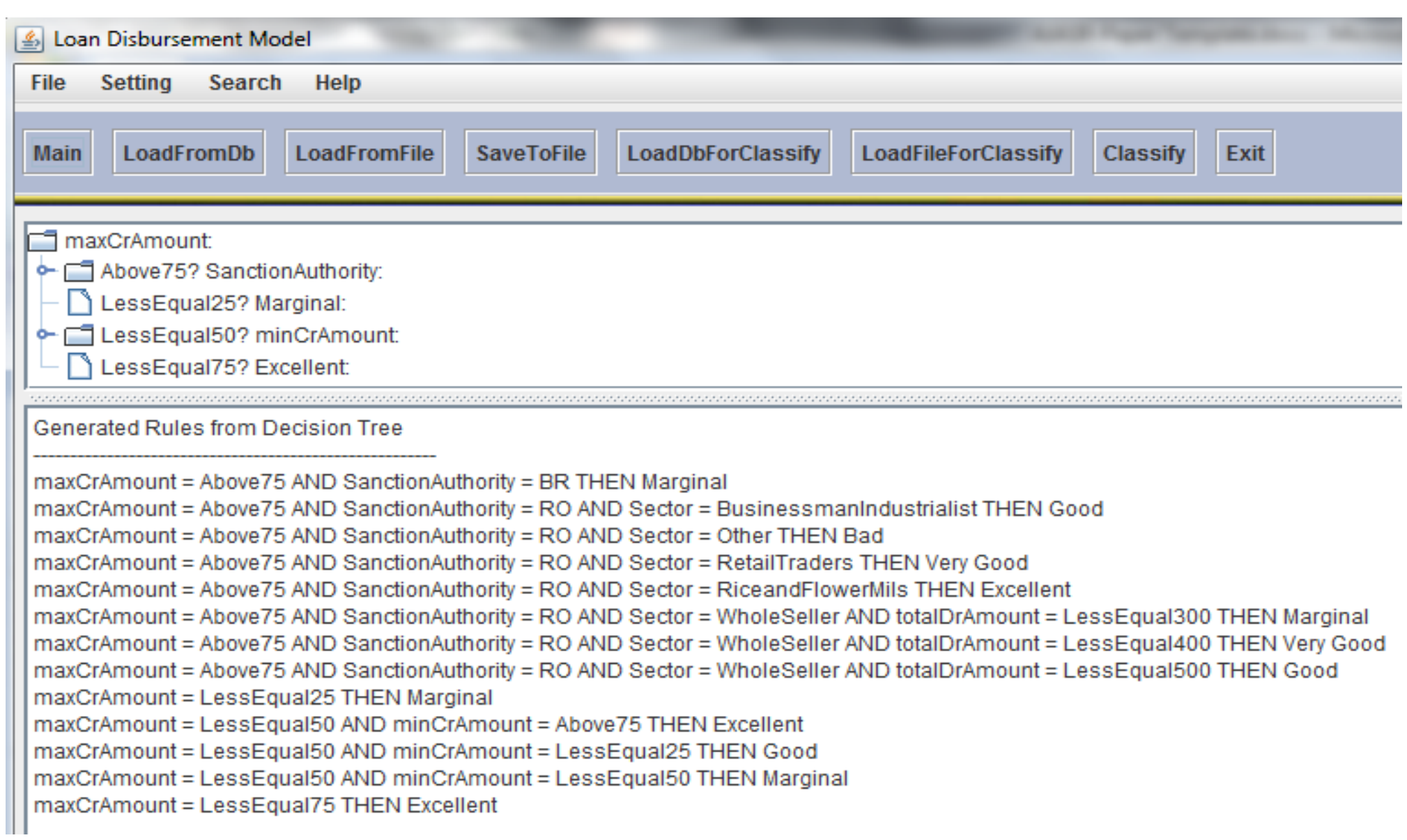

Figure 6. Generated rules from pruned tree

## 5. RESULT AND DISCUSSIONS

### 5.1. Testing

The training data were collected from transactional database not later to 30/06/2014 and not prior to 01/01/2013. The model is tested with some samples data prior to 01/01/2013 and later to 30/06/2014 from the same database. The model also tested with data collected from a nearby bank branch. The initial findings show some interesting results.

### 5.2. Comparison with Weka result

To test the effectiveness of the model, the model is tested with the Weka software and found that the outputs from them are almost same. The output of weka result from a nearby branch data is shown in "Figure 7". Comparing of "Figure 6" and "Figure 7", it can be said that the model is accurate enough to predict prospective business customers in retailing commercial bank.

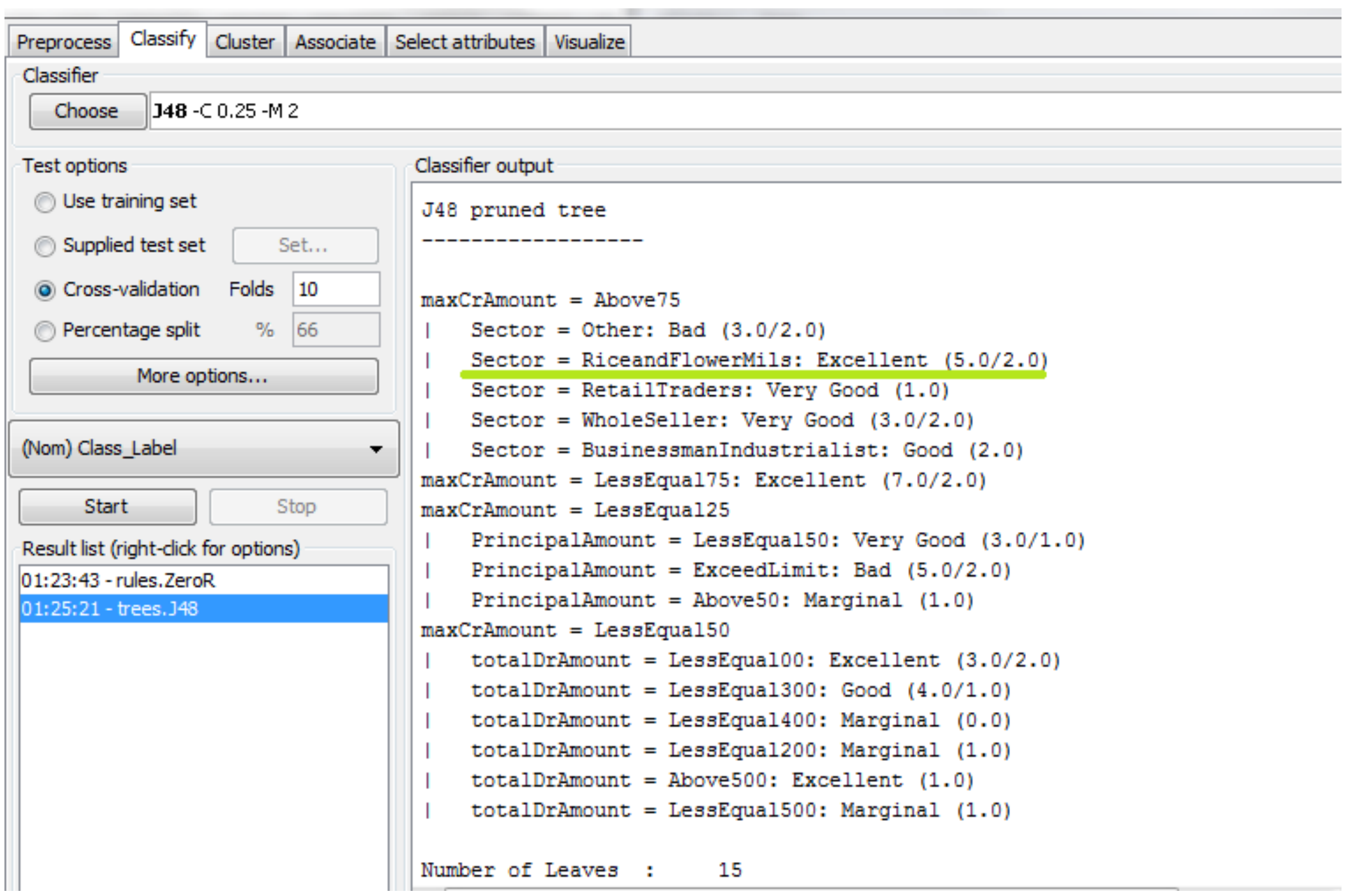

Figure 7. Output of Weka Software from a nearby branch data

### 5.3. Data Mining Finding

The initial studies unveiled a number of relationships between variables that justify further discussion and analysis. Sectors are business area in where the loan disbursed. Our finding indicates RiceAndFlowerMis are excellent (best) to disburse loan in the researched area. RetailTraders and WholeSeller sectors are also very good for lending.

## 6. CONCLUSIONS

We have discussed and implemented the procedure of a data mining task for prospective business sectors analysis in retail banking. We have predefined the optimal number of clusters (Excellent, Very Good, Good, Marginal and Bad) and used classification methods (Decision Tree) to predict prospective business sectors based on existing customer transactional behavioural data. The initial findings show some interesting results. The field test conducted and found that the data mining prediction model for prospective business sectors for disbursement of loan is very accurate and the target-oriented campaign is very effective. The results show that most prospective customers are in Rice and Flower Mils (RiceAndFlowerMils) sectors; and Retail Traders (RetailTraders) and Whole Seller (WholeSeller) are also very good.

The main focus of the paper was on written attributes of business customer but customer have many unwritten behaviour attributes. So the topic exclude some important unwritten attributes such characters, conditions and collaterals. The model is developed based on data mining method of decision tree classification as an initial step. We have a plan to extend the model for other data mining methods in future.

## ACKNOWLEDGEMENTS

We have pleased to all managers, assistant managers, loan concerned offices of branches and loan disbursement authorities of regional office, Tangail for their cooperative understanding, help and support during completion of our work. We also wish to thank the developers of Java and Weka.

## AUTHORS

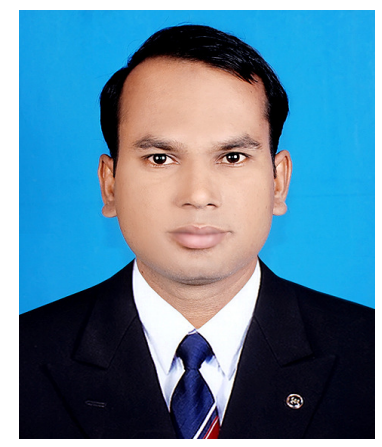

Md. Rafiqul Islam was born in 1986 in Bangladesh. He has received his B.Sc (Engg) in Information and Communication Technology from Mawlana Bhashani Science and Technology University, Tangail, Bangladesh in 2010 and is currently pursuing M.Sc (Engg) in Information and Communication Technology at the same university. He is also serving as an Assistant Programmer (Assistant Director) at Bangladesh Bank (Central Bank of Bangladesh), Dhaka, Bangladesh. Formerly, he had served in Janata Bank Limited and in software development companies (Documenta, Horoppa InfoTech) in Bangladesh. His technical areas of interest include Data Mining, Artificial Intelligent, Image Processing, Software Security, Application Development and Algorithms.

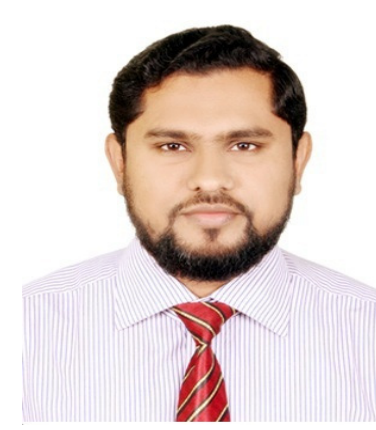

Md. Ahsan Habib was born in 1982 in Bangladesh. He is one of the PhD research fellow in Green Networking Research Group in CSE Department of Dhaka University, Bangladesh. Mr. Habib received BSc in Computer Science and Information Technology and MSc in Computer Science and Engineering from Islamic University of Technology, Dhaka, Bangladesh in 2003 and 2012, respectively. He is serving as an Associate Professor in the Department of Information and Communication Technology in Mawlana Bhashani Science and Technology University, Tangail, Bangladesh. He has been teaching a good number of courses related to wireless communications, wireless sensor network, bioinformatics, data mining and warehouse, cyber security etc. to graduate and undergraduate students in reputed universities. Formerly, he had served in Hajee Mohammad Danesh Science and Technology University Dinajpur, Bangladesh. He is a member of IEB, IEEE, Bangladesh Physical Society, Bangladesh Computer Society and Lions Club International.